\documentclass[amsmath,amssymb,apl,preprint]{revtex4}
\usepackage{graphicx}
\usepackage{bm}

\begin{document}


\title{Characterization of dipolar current-based metamaterials showing a single resonance in the terahertz spectrum}

\author{C. Rizza}
\affiliation{Dipartimento di Ingegneria Elettrica e dell'Informazione, Universit\`{a} dell'Aquila 67100, Monteluco di Roio - Italy}
\author{E. Palange}
\email{elia.palange@univaq.it} \affiliation{Dipartimento di Ingegneria Elettrica e dell'Informazione, Universit\`{a} dell'Aquila 67100, Monteluco di
Roio - Italy}
\author{P. Carelli}
\affiliation{Dipartimento di Ingegneria Elettrica e dell'Informazione, Universit\`{a} dell'Aquila 67100, Monteluco di Roio - Italy}
\author{M. Ortolani}
\affiliation{Istituto di Fotonica e Nanotecnologie, IFN-CNR, Via Cineto Romano 42, 00131 Roma - Italy}
\author{O. Limaj}
\affiliation{CNR-INFM Coherentia and Dipartimento di Fisica, Universit\`{a} di Roma "La Sapienza" P.le Aldo Moro, 2 - Italy}
\author{A. Nucara}
\affiliation{CNR-INFM Coherentia and Dipartimento di Fisica, Universit\`{a} di Roma  "La Sapienza" P.le Aldo Moro, 2 - Italy}
\author{S. Lupi}
\affiliation{CNR-INFM Coherentia and Dipartimento di Fisica, Universit\`{a} di Roma  "La Sapienza" P.le Aldo Moro, 2 - Italy}

\date{\today}

\begin{abstract}

We will report on the electromagnetic response due to induced dipolar currents in metamateri-als of $2$-dimensional array of metallic elements. Used
as frequency selectors, the metamaterial transmittance presents a single resonance in the region from $1$ to $8$ THz that can be easily selected and
scaled maintaining unaltered the quality factor by choosing the size and shape of the planar metallic element and exploiting the scalability
properties of the Maxwell equations. Basing on these studies, we have designed and tested a series of simple and inexpensive frequency selective
metamaterials fabricated by using lithographic processes.

\end{abstract}

\maketitle
\begin{figure}[h]
\begin{center}
\includegraphics[width=0.8\linewidth]{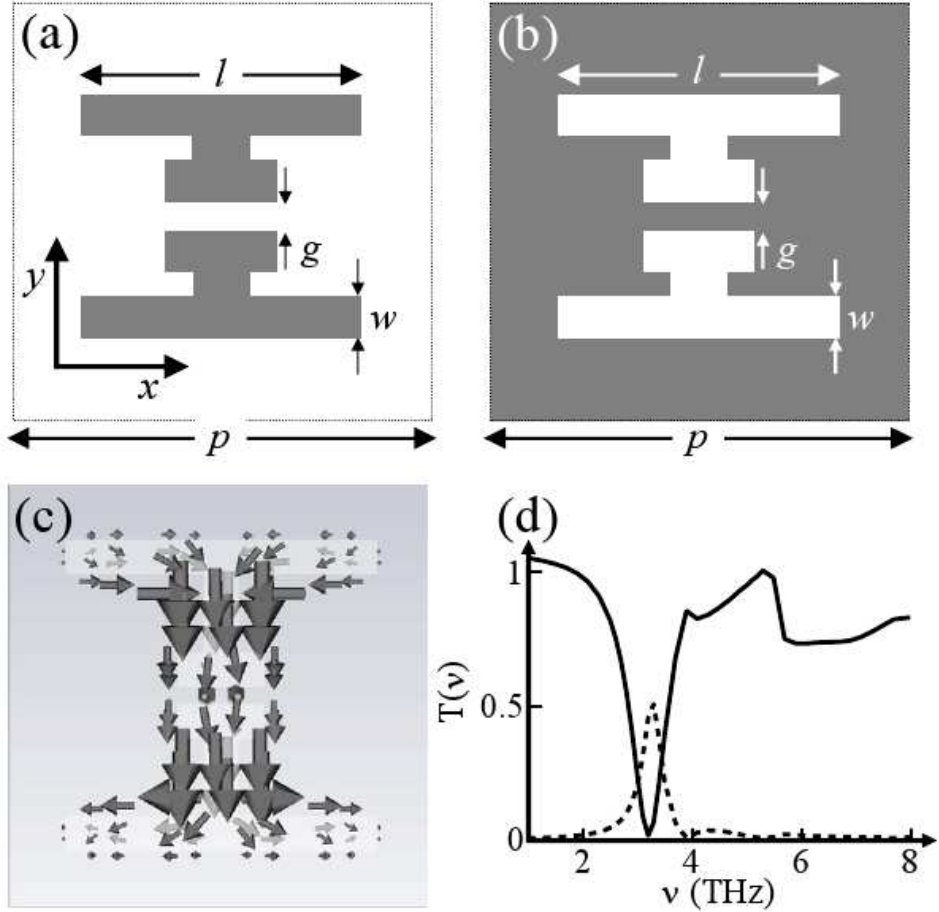}
\caption{Geometry for the direct (a) and complementary (b) planar metallic element: $l$ is the length of the horizontal stripe, $w$ the width of the
stripes, $g$ the gap distance and $p$ the $2$D-array lattice constant. (c) Plot of the induced surface dipolar current density distribution for the
element of panel (a). The input polarization is along the $y$-axis. (d) Transmittance of the direct (solid line $y$-axis polarization) and
complementary MM (dashed line $x$-axis polarization). For (c) and (d) all dimensions have been chosen to obtain a resonance at $3$ THz.} \label{fig1}
\end{center}
\end{figure}

Scientific and technological research in the terahertz (THz) frequency region of the electromagnetic (e.m.) spectrum can get benefits from the
development of optical components and devices based on MetaMaterial (MM) structures that find applications in chemical and astronomic spectroscopy,
material science, medical and biological analysis and imaging, communication networks, defence and security screening \cite{Siegel,Ferguson}. In
order to provide a widely usable THz technology, the research on the MM optical components has been mainly oriented in developing selective frequency
filters, advanced focusing systems, polarization controllers and rotators \cite{Chen1,Bingham,Wu,Maier,Flomanek,Agrawal,Peralta}. On this respect,
MMs mainly overcome the absence of a natural response of materials in the THz region. In particular, suitable designed planar metallic elements can
be arranged in $2$-dimensional ($2$D) periodic arrays on dielectric or semiconductor hosting substrates forming MM optical systems. MMs show
transmittance and reflectance response to an incident e.m. radiation that can be related by the Fresnel relations to an effective dielectric
permittivity and/or magnetic permeability that, in turn, depend on the frequency and polarization status of the e.m. radiation as well as on the
shape and size of the planar metallic element. In general, MMs present a bianisotropic behaviour since the induced electric displacement and magnetic
fields are coupled through magneto-optical permittivities \cite{Padilla1,Padilla3}. A purely electric response can be obtained in $2$D-arrays of
Double Split Ring Resonators (DSRRs) whose transmittance in the THz region has been studied in several cases \cite{Padilla2}. Regardless of the
particular shape of DSRRs, two main resonances are typically observed in their transmittance response for an input polarization of the e.m. radiation
parallel to the split gap characterizing the DSRR geometry. The first, at the lower frequencies, is a narrow inductive-capacitive ($LC$) resonance
caused by charge accumulation at the split gap and circulating currents in the DSRR loops. At higher frequencies, a second broad absorption feature
is observed that is instead associated to the presence of dipolar currents. This dipole resonance is the only one that survives (even if red-shifted
and more broadened) when the input polarization is set perpendicular to the split gap. This result demonstrates that the dipolar-related resonances
depend on the interactions between the components of the planar metallic elements parallel to the polarization status of the incident e.m. radiation
\cite{Gotschy,Ditlbacher,Rockstuhl,Azad}. By exploiting the Babinet's principle, it is also possible to investigate the behaviour of
\emph{complementary} DSRR-based MMs in which, respect to \emph{direct} DSRRs, metal replaces voids and vice versa \cite{Chen2,Jackson}. In this case,
a minimum in the transmittance observed in direct DSRRs generally corresponds to a maximum in the transmittance in the complementary ones. It is
worth noting that used as frequency filters, both direct and complementary DSRRs exhibit poor off-resonance selectivity especially at the higher
frequencies, detrimental for the resulting quality factors. For example, in the case of the presence of both $LC$- and dipole-related resonances, the
observed transmittance features are comparable in strength. As a consequence, for particular DSRR shapes and sizes these physical mechanisms can
produce close enough frequency resonances strongly reducing the DSRR selectivity properties. An interesting solution that partly overcomes this
problem is to change the dimension of the planar metallic element in such a way to set far away the $LC$- and dipole-current-related resonances
\cite{Azad}. However, the transmittance spectrum continues to present two distinct resonances that cannot be easily selected because both depend on
the dimension and reciprocal distance between neighbouring planar metallic elements.

The aim of the present paper is to investigate the properties of a frequency-selective MM structure of $2$D-array of planar metallic elements whose
transmittance response to an input e.m. radiation presents a single resonance in the frequency region from $1$ to $8$ THz that is related only to the
induced dipolar currents. By choosing an appropriate size of the planar metallic element and $2$D-array lattice period, the MM transmittance
resonance can be selected, whereas the scaling of the structure allows to shift the resonance maintaining unaltered the quality factor. The basic
element of the frequency-selective MM has been a modified version of the DSRR of Ref.\cite{Chen1} in which it is easy to suppress any net circulating
current by removing the vertical stripes as shown in Fig.$1$(a). The complementary element of Fig.$1$(a) is reported in Fig.$1$(b). The surface
current density distribution and the transmittance curves of these MMs have been determined by numerically solving in the $3$D spatial coordinates
and in the time domain the Maxwell equations using the CST Microwave Study commercial software. In Fig.$1$(c) is shown the plot of the absolute value
of the surface current density distribution in the element of Fig.$1$(a) for a resonant frequency at $3$ THz for a polarization of the input
radiation along the $y$-axis (i.e. parallel to the split gap) that makes evident the dipolar nature of the induced currents. For both the elements,
direct and complementary, the typical transmittance curves centred at $3$ THz have been calculated and reported in Fig.$1$(d). In order to take into
account the reflection from the substrate surfaces, the transmittance, $T=I_{sam}/I_{ref}$, at each frequency has been determined by calculating the
ratio between the intensity $I_{sam}$ obtained with the planar circuit $2$D-array deposited upon the substrate divided by the intensity $I_{ref}$ of
the substrate alone. The transmittance curves for both the direct and complementary MMs present a single resonance: respect to the direct DSRR the
$LC$ transmittance feature is absent and therefore any interactions between dipolar and $LC$ resonances are suppressed \cite{Chen1,Padilla3}. This
provides a strong benefit in the filter response of the proposed frequency-selective MMs. According to simulations, the frequency-selective quality
factor of direct MMs can be increased approaching the value of $30$ by varying the split gap distance, the stripe width and the lattice constant of
the $2$D-array. An enhancement of the quality factor is always associated to an increase (decrease) of the value of the transmittance minimum
(maximum) for direct (complementary) MMs. In addition, the transmittance feature for complementary MMs is poorer if compared to that one for direct
MMs, since the Babinet's principle completely holds only when the film thickness of the planar metallic element approaches zero within the further
approximation to use a perfect conductive material \cite{Chen2,Jackson}. Our choice in designing the experimentally tested MMs has been to study the
direct and complementary MMs having same sizes and for this we have limited the achievable quality factor for the direct MMs in order to have a
detectable intensity of the radiation passing through the complementary MMs. We note that circulating currents can be eliminated also removing the
two horizontal stripes in the DSRR of Ref.\cite{Chen1}. However, in this case simulations show that three strongly interacting dipoles are induced in
the planar circuit (one \emph{per} each surviving metallic structures) lowering the quality factor of the resulting transmittance curves. This agrees
with the fact that for the dipole induced transmittance the quality factor mainly depends on the radiative resistance that, in turn, scales with the
distance between neighbouring dipoles \cite{Mey}. The tailoring of the frequency-selective MMs at any desired frequency can be obtained by using the
scalability properties of the Maxwell equations that are invariant under the following linear scale transformations: ${\bf E}'=\alpha {\bf E}$, ${\bf
B}'=\alpha {\bf B}$, ${\bf r}'=\alpha {\bf r}$, where the frequency and the conductivity become, respectively: $\nu'=\nu/\alpha$,
$\sigma'=\sigma/\alpha$. ${\bf E}$, {\bf B} are the electric and magnetic fields, ${\bf r}$ the position vector and $\alpha$ the scale factor. The
figure of merit of the frequency selective MM, i.e. its resonance frequency and quality factor, depends on the size of the planar element forming the
MM once its shape has been defined. Therefore, when the size of all the constituents of the planar metallic element are scaled by $\alpha$, the
resonance frequency is varied by the same quantity preserving, at the same time, the quality factor defined as $Q=\nu_0/\Delta \nu$, where $\Delta
\nu$ is the full-width at half-maximum of the transmittance $T(\nu)$ and $\nu_0$ the resonant frequency. This result is in agreement with the fact
that in the THz region the MM response is not affected by plasma resonances that limit the linearity of the size scaling \cite{Klein}. We remark that
in the scale transformation also the conductivity $\sigma$ should be scaled by $\alpha$. Indeed, $\sigma$ is related to the skin depth effect that,
in turn, scales through the same factor. However, in real situations, $\sigma$ cannot be varied as it is a property of the metal used to fabricate
the planar elements. As a consequence, the skin depth scales as $1/\sqrt{\alpha}$. We used the scale transformations, to perform a detailed numerical
analysis for both direct and complementary MMs. Referring to Fig.$1$(a), we have chosen as starting point a MM-based filter with a resonance centred
at $2$ THz with the following geometrical dimensions: a horizontal and vertical outer size $l=25.4 \mu m$, a gap $g=1.4 \mu m$, a metal strip width
$w=2.9 \mu m$ and a $2$D-array lattice period $p=35.7 \mu m$. The factor $\alpha$ for which these geometric sizes have to be scaled to obtain
transmittance resonances at $3$, $4$, and $5$ THz are therefore $1.5$, $2$, and $2.5$, respectively. To simulate the transmittance in the THz range,
an array of $220 nm$ thick aluminium planar elements with the previous geometric dimensions has been located on the $(x,y)$ plane of a $300  \mu m$
thick substrate with a dielectric constant $11.6$ equal to that one of silicon and $z$ was chosen as the propagating direction of the THz radiation.

Basing on the results of the numerical simulations, experimental investigations have been performed by fabricating $4$ direct and $4$ complementary
frequency selective MMs scaled to obtain transmittance resonances centred at $2$, $3$, $4$ and $5$ THz, through a single lithographic process. By
using a double-side polished Si wafer substrate, the lift-off process was based on $UV3$ (a positive tone chemical amplified resist) patterned by
electron beam lithography direct writing. Each MM $2$D-array fills an area of $5 \times 5 mm^2$ and a deposited $220nm$ thick layer of aluminium is
lifted-off by soaking it in acetone. The THz transmittance of all MMs in the $1-8$ THz range were measured at room temperature through an evacuated
Fourier Transform Bruker $IF66v$ interferometer, using radiation from a mercury source that is linearly polarized both parallel or perpendicular to
the split gap direction. The sample was positioned in the focus of an $f/4$ parabolic mirror and the transmitted beam was collected by a twin $f/4$
mirror. The collimated beam after the second mirror was then focused into a far-infrared pyroelectric detector.
\begin{figure}[h]
\begin{center}
\includegraphics[width=0.5\linewidth]{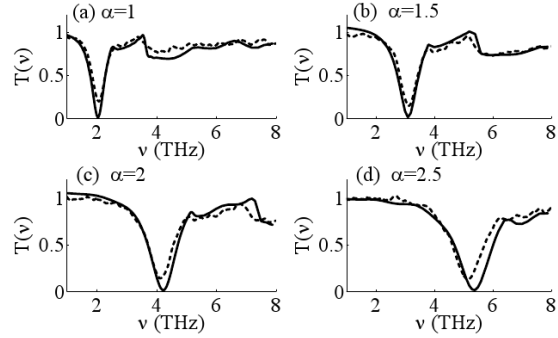}
\caption{Transmittance for the direct MMs. Dashed and solid lines refer to the experimental and calculated data, respectively. The resonant frequency
has varied by using the scale factor $\alpha$ as outlined in the text. The input THz radiation is polarized parallel to the split gap.} \label{fig2}
\end{center}
\end{figure}
In Fig.$2$(a-d) we report the transmittance curves of the direct MMs (dashed line) with an input beam polarized parallel to the split gap: each MM
transmittance curve is characterised by a single resonance. For all the frequency selective MMs, the resonances are about $5\%$ blue-shifted respect
to the expected values. This slight difference can be related to some systematic error in the etching procedures during the sample fabrication
process: the overall resulting planar element dimension is smaller than that one designed by electron beam lithography. Direct scrutiny of the planar
circuit dimensions performed by atomic force microscopy confirms this assumption. However, this result suggests that the frequency-selective MMs have
a great sensibility to structure size variations and this dependence can be useful for fine frequency tuning of the MM resonance. In Fig.$2$(a-d),
the solid lines are the calculated transmittance as obtained by performing the numerical simulations on $5\%$ in dimension reduced planar elements.
Fig.$3$(a-d) reports the measured transmittance curves (dashed lines) for the complementary frequency selective MMs and the corresponding calculated
curves (solid line).
\begin{figure}[h]
\begin{center}
\includegraphics[width=0.8\linewidth]{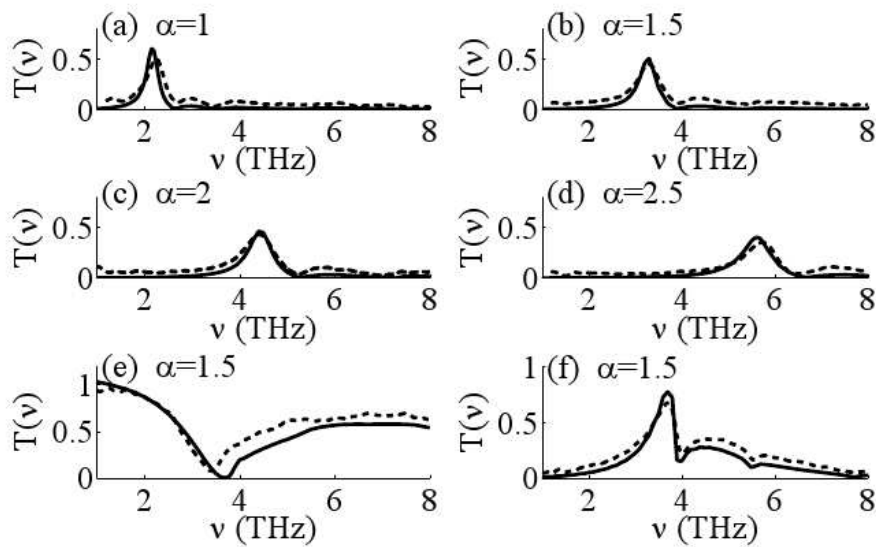}
\caption{Transmittance for the complementary MMs (a-d). Dashed and solid lines refer to the experimental and simulated data, respectively. The
resonant frequency has varied by using the scale factor $\alpha$ as outlined in the text. The input THz radiation is polarized perpendicular to the
split gap. (e) and (f) transmittance at $3$ THz of the original and complementary MMs by using a polarization of the incident radiation perpendicular
and parallel to the split gap, respectively.} \label{fig3}
\end{center}
\end{figure}
Fig.$2$(a-d) and $3$(a-d) show that the calculated transmittance curves are in good agreement with the experimental findings. Regarding the quality
factor, as stated above, its value is strictly independent on the resonance frequency if the conductivity and the metal thickness have been properly
scaled by $\alpha$. However, this seems to be a small correction since both the calculated and experimental quality factors of the transmittance of
Fig.$2$(a-d) and $3$(a-d) are constant with respect to the variation of the resonant frequencies. In particular, the experimental values of the
quality factor can be evaluated equal to about $4.5$ and $6.7$ for the direct and complementary MMs, respectively, close to the calculated values
$4.2$ and $7$. The transmittance curves of MMs having a resonance at $3$ THz obtained by using a polarized radiation perpendicular (parallel) to the
split gap for the direct (complementary) MMs are reported in Fig.$2$ (e) and (f), respectively. We note a decrease in the quality factor together
with a blue-shift of the resonant frequencies, the latter confirming its dependence on the distribution of the dipolar currents and their
interactions between neighbouring metallic planar elements \cite{Gotschy}. In conclusion, we have numerically and experimentally shown that it is
possible to realise a frequency selective MMs based on an array of suitable planar elements that can be designed to present a single resonance in the
transmittance over the spectral region ranging from $1$ to $8$ THz with quality factors comparable with those ones reported in literature. The MM
frequency response to the THz input radiation depends on the induced dipolar currents since net circulating currents are not allowed in the planar
metallic element. The frequency selective MM resonance can be centred at any frequency by varying the dimensions of the planar elements using a
geometric scale factor directly deduced by the scalability properties of the Maxwell equations. We believe that the overall behaviours of the
investigated MMs that require for their fabrication only few steps of lithographic process, represent an improvement in the development of
inexpensive THz filters having characteristics similar to those ones of coloured optical filters in the visible.
\newline
The authors acknowledge D. Doddi and A. Priante for their support in the metamaterial design and fabrication.

\end{document}